\documentclass[12pt]{iopart}

\usepackage{color}
\usepackage{cite}
\usepackage{amssymb}
\usepackage{graphicx}   
\usepackage{subfigure}
\usepackage{bm}

\usepackage{gensymb}
\usepackage{hyperref}
\hypersetup{
    colorlinks=true,
    linkcolor=blue,
    filecolor=magenta,      
    urlcolor=cyan,
    }

\begin{document}

%
\title{Dislocation-CNT Interactions in Aluminium at the Atomic Level}

\author{Samaneh Nasiri$^1$, Michael Zaiser $^1$ }
\address{$^1$WW8-Materials Simulation, Department of Materials Science, Friedrich-Alexander Universit\"at Erlangen-N\"urnberg, 90762 F\"urth, Germany} 
\ead{samaneh.nasiri@fau.de}

\begin{abstract}
The interaction between edge dislocations and carbon nanotubes (CNTs) embedded into an Al matrix is investigated. Both pristine and Ni-coated CNTs are considered. It is demonstrated that the embedded CNTs lead to a flow stress contribution equivalent to that of an array of non-shearable inclusions that are by-passed via the Orowan mechanism. The strain hardening mechanism associated with multiple dislocation passes shows, however, characteristic differences from the Orowan picture that indicate that the embedded CNTs behave analogous to embedded voids. 
\end{abstract}

\section{Introduction}
\label{sec:1}

Carbon nano-particles (CNP), in particular carbon nanotubes (CNT) and Graphene (GP) sheets with their excellent mechanical properties \cite{Lee2008-science, Popov2000-PRB, Yu2000-science, nasiri2016-AIMSmaterialsscience} are considered as promising candidates for fillers in composites where the matrix is a lightweight metal like Al and Mg. Such composites might have applications in the aeronautical, electrical and automotive industries. The benefits, in view of mechanical properties, of embedding well-dispersed carbon nanoparticles into metals are potentially huge, for discussion see e.g. \cite{George2005-scripta, Li2009-cops.sci.technol, Bakshi2010-Int.mater.rev, Nasiri2019-EPJ}. They derive on the one hand from classical composite mechanics – strong reinforcing fibers and platelets will carry part of the load and increase effective elastic moduli – on the other hand from the interaction of nanoparticles with matrix defects and from their interference with deformation and failure mechanisms in pristine metals: Strong embedded nanoparticles can potentially bridge incipient cracks and impose strong constraints on dislocation motion, thus potentially increasing yield strength, fracture strength, and possibly also toughness. Added benefits can be drawn from the nanoscale dimensions of the reinforcing particles in terms of small spacing (high Orowan stresses), high specific surface area and very high aspect ratio of the tubular or flake- like particles, resulting in efficient crack bridging and high creep resistance as dislocations cannot easily climb around the embedded particles \cite{Chen2015-CST,Chen2017-Acta,Chen2019-Jalcom, Park2018-JMST,Silvestre2014-comscitech,Choi2016-JCOMB,Xiang2017-commatsci, Li2010-compos.sci.technol, So2016-eml, Lahiri2009-MSE:A}.

Atomistic simulation can play an important role for understanding streng\-thening mechanisms in CNP reinforced metals, and in particular the interactions of CNPs with dislocations. This can be envisaged in analogy with interactions of dislocations with other types of embedded particles such as precipitates, for which a vast body of atomistic simulation studies exists, exploring aspects such as the role of matrix/precipitate interface, precipitate stiffness, size, shape, orientation of precipitate and dislocation type \cite{Proville2010-Acta, Terentyev2008-J.Phys.condens.Matt, Terentyev2011-J.nuc.mat, Prakash2015-Acta, vaid2019-Materialia, Bonny2019-ScientificReports}. An extensive review of atomistic simulation of dislocation-obstacle interactions considering different parameters in various metals can be found e.g. in \cite{Bacon2009-disinsol}.

Regarding the interaction of dislocations in metals with embedded CNPs, there exist few simulation studies. Kim et al. \cite{kim2013-Nature} complement their experimental investigation of layered metal-graphene nanolayer composites with molecular dynamics (MD) simulations of the interaction of a dislocation with an embedded graphene layer and find that dislocations get arrested by elastic interactions with the embedded graphene before they reach the graphene-metal interface. A similar dislocation arrest mechanism was deduced by Chang et al. \cite{Chang2013-Philos.Mag.Lett} from MD simulations of indentation tests on Ni-GP nanolayers. Liu et al. \cite{Liu2016-carbon} use MD to study the shearing of Cu-G multilayers with (111) and (100) stacking and dislocation-free Cu matrix, and shear conducted along [100] (100) and [112] (111) directions in such a manner that the graphene layers intersect the plane of shear. In these cases a significant strengthening was found as compared to pristine Cu; this was attributed to the dual confining effect of the Cu matrix on the GP filler (prevention of wrinkling) and of the GP sheets on the Cu matrix (blocking of dislocations). 

In this study we focus on reinforcement by carbon nanotubes and present atomistic simulations of edge dislocations in Al interacting with pristine and nickel coated CNTs (henceforth referred to as NiCNTs) of different size and inter-particle spacing. The rationale of considering NiCNTs for Al matrix reinforcement has been explored elsewhere: Ni coating of CNTs very significantly improves interfacial adhesion between Al matrix and reinforcing CNTs, thus improving load transfer and preventing nanotube pull-out \cite{Nasiri2019-EPJ}. Here we investigate how the modified interface affects the interaction between embedded CNTs and matrix dislocations.

\section{Methods and Results}
\label{sec:2}
The atomistic simulations presented in this work were performed with the molecular dynamics code LAMMPS \cite{LAMMPS} (version 16 Mar 2018). Atomsk package \cite{Atomsk}, and VNL builder in QuantumATK \cite{schneider2017-iop,Stradi2017-iop} were used to create the initial atomic configurations. Data visualization and post-processing were obtained with the help of the Open Visualization Tool (OVITO) \cite{ovito}. Common neighbour analysis (CNA) \cite{Faken1994-comput.mat.sci, Honeycutt1987-J.Phys.Chem} was used to analyze the crystallographic structure around atoms and to identify defects and dislocations in the Al matrix.  

\subsection{Interatomic potentials}
\label{sec:2.1}
To describe atomic interactions, we use the standard AIREBO potential \cite{Brenner2002-J.Phys.Condens.Matter,stuart2000-J.Chem.Phys.} for carbon-carbon, and the EAM/alloy potentials of Mishin \cite{mishin2004-acta} for metal-metal interactions. Interfacial interactions between CNTs and metals are described by using Morse potentials for Al-C and Ni-C interfaces which have been parametrized using DFT calculations and fitted to represent physisorbed Al/GP and Ni/GP interfaces \cite{Nasiri2019-EPJ}. The energy function in the Morse potential has the form $E_{\mathrm{M}} = D_{\mathrm{M}} (\exp[-2\alpha (r - r_0)] - 2 \exp[-\alpha (r - r_0)])$, where $r$ is the bond distance, $r_0$  an equilibrium bond distance, $D_\mathrm{M}$ is the well depth, and $\alpha$ controls the stiffness of the potential. The obtained parameters for the Al-C Morse potential are $D_\mathrm{M}$ = 0.004 eV, $r_0$ = 4.52 $\mathrm{\AA}$ and $\alpha$ = 1.0 $\mathrm{{\AA}^{-1}}$ and for the Ni-C Morse potential $D_\mathrm{M}$ = 0.0048 eV, $r_0$ = 4.18 $\mathrm{\AA}$ and $\alpha$ = 1.15 $\mathrm{{\AA}^{-1}}$. 
 
\subsection{Sample Preparation}
\label{sec:2.2}

\begin{figure}
	\centering
		\includegraphics[width=.7\textwidth]{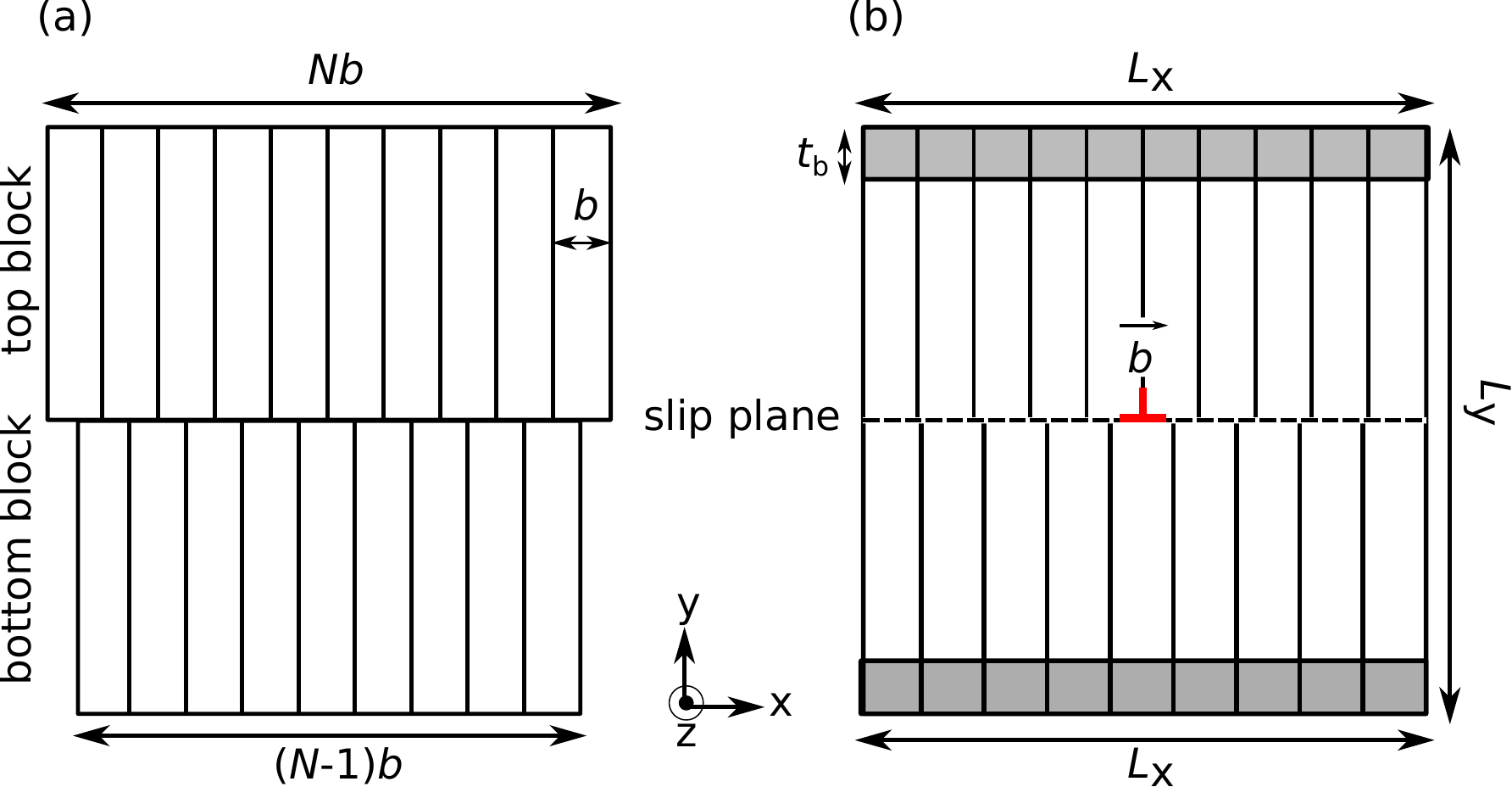} 
	\caption[Schematic illustration of construction of an edge dislocation]{Schematic illustration of construction of an edge dislocation, $N$ is the number of lattice planes and $b$ is the Burgers vector magnitude, (a) two crystal blocks of Al are placed on top of each other. The top block has one lattice plane more than the bottom one. (b) The top block is subjected to affine compression in $x$ direction with a total length reduction of $b/2$, and the lower block to affine elongation by $b/2$, such that both have the same length in $x$ direction. The blocks are then merged along the (-1-1-1) plane. Periodic boundary conditions are imposed in $z$ direction (line direction of the dislocation) and in $x$ direction (glide direction).}
	\label{fig:dis0}     
\end{figure}

We use the Atomsk package \cite{Atomsk} to construct an aluminum block in the form of a parallelepiped containing an edge dislocation, following the procedure described in Ref. \cite{Bacon2009-disinsol}. The edges of the Al block are aligned with the axes of a Cartesian coordinate system where the $z$  axis points in the [-1-12] direction, the $y$ axis in the [-1-1-1] direction, and the $x$ axis in the [-110] direction of the fcc crystal lattice. An edge dislocation of Burgers vector $\vec{b} =b/\sqrt{2}[-110]$ is created by cutting the block along the (-1-1-1) plane (the $xz$ plane of the coordinate system) and adding one (-110) lattice plane in excess to the top half (Fig. \ref{fig:dis0} (a)). The upper half is then subjected to affine compression in $x$ direction with a total length reduction of $b/2$, and the lower half to affine elongation by $b/2$, such that both have the same length in $x$ direction. The blocks are then merged along the (-1-1-1) plane. Periodic boundary conditions are imposed in $z$ direction (line direction of the dislocation) and in $x$ direction (glide direction). After constructing the initial simulation box, the system is relaxed using the fast inertial relaxation engine (FIRE) algorithm \cite{FIRE}. The system is accepted as relaxed when both the change of energy between successive iterations divided by the energy magnitude is less than or equal to the tolerance ($10^{\rm -8}$) and the 2-norm (length) of the global force vector is less than the threshold value of $10^{\rm -8}$ eV/$\mathrm{\AA}$ as in \cite{vaid2019-Materialia}. The shaded regions in Fig. \ref{fig:dis0} show the boundary layers. Atoms in the boundary layers of thickness $t_{\rm b}$ = 14 $\mathrm{\AA}$ are considered completely rigid during the relaxation by setting all forces on these atoms to zero. After the relaxation, an edge dislocation of Burgers vector $\vec{b} =b/\sqrt{2}[-110]$ is present in the Al matrix. 

\begin{figure}
	\centering
		\includegraphics[width=.45\textwidth]{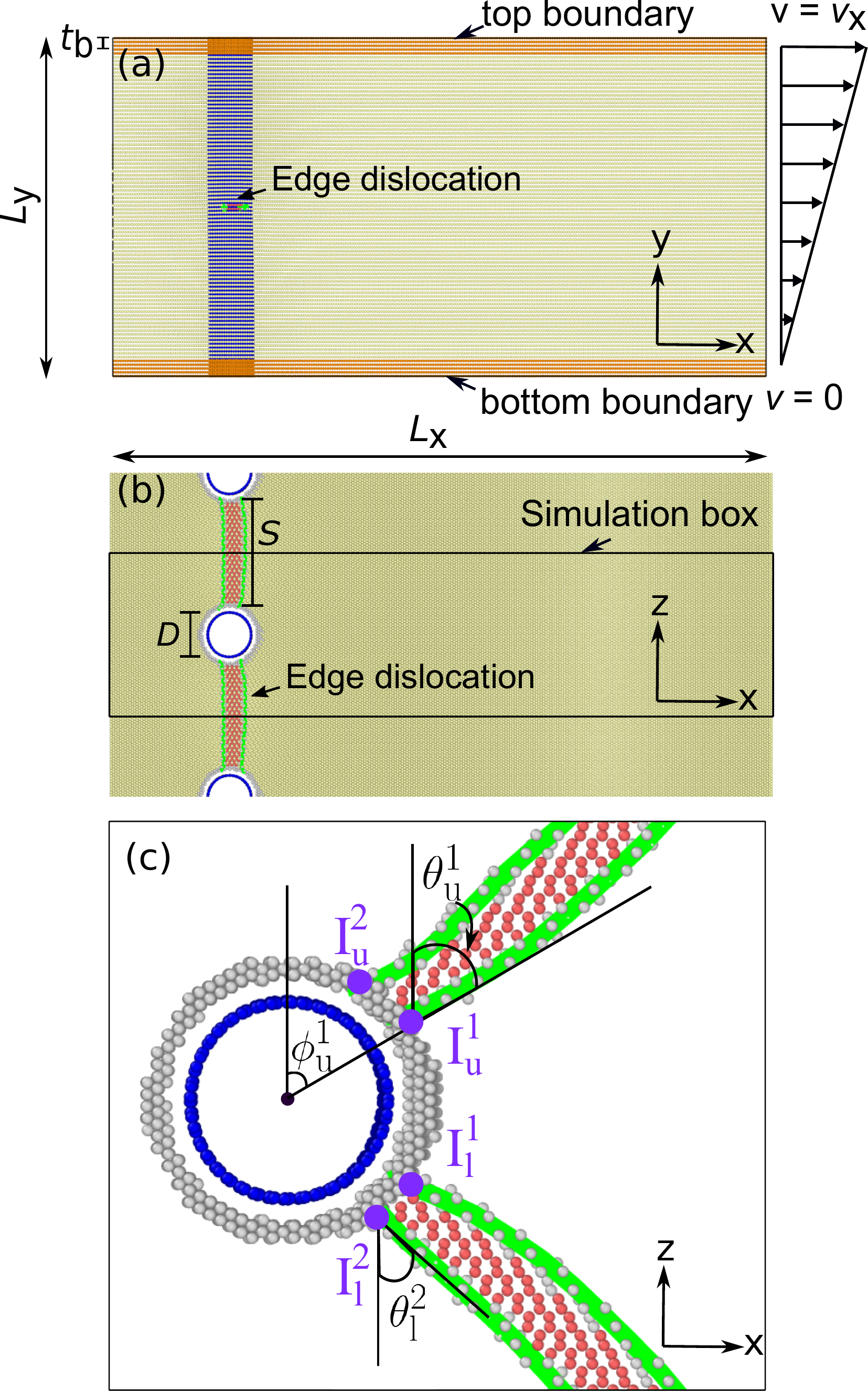} 
	\caption[Edge dislocation in Al/CNT]{Initial configuration of a dislocation interacting with a CNT after relaxation and before loading,  (a) side view and (b) top view on the dislocation slip plane.  Al atoms whose nearest neighbours form fcc and hcp environments are depicted in yellow and red color, respectively. Carbon atoms of the pristine CNT(10,10) are shown in dark blue. Al atoms at the CNT interface have surroundings of reduced symmetry and are shown in light grey color. Periodic boundary conditions apply in $x$ and $z$ directions.  The top and bottom layers of atoms of thickness $t_\mathrm{b}$ = 14 $\mathrm{\AA}$ are constrained, these atoms are coloured in orange. During relaxation the dislocation splits into two partial dislocations separated by a stacking fault and aligns with the CNT and its periodic images. The solid green lines indicate partial dislocation lines. (c) Detail of a dislocation under load close to the CNT, showing the intersection points (in violet color) of the leading and trailing partials of the upper and lower arms.} \label{fig:1}     
\end{figure}

The next step is to embed a single walled carbon nanotube (SWCNT) into the Al block in such a manner that it is threading the Al block from top to bottom along the $y$ axis. We consider pristine as well as Ni coated CNTs. The procedure for Ni coating has been discussed in detail elsewhere \cite{Nasiri2019-EPJ}. To equilibrate Ni atoms on the CNT surface and allow for the possible formation of a continuous Ni coating  \cite{Li1997-JJAP,Kong2002-sct}, a layer of Ni atoms of thickness 7.0 $\mathrm{\AA}$ around a CNT(10,10) with a diameter of 14 $\mathrm{\AA}$  is annealed at 2300K (i.e., above the melting temperature of Ni) for 20 ps in the $NpT$ ensemble at zero pressure, and then quenched to 0.1 K at a rate of 10 K/ps. During this anneal-quench cycle, the CNT atoms are kept fixed. To embed a pristine or Ni-coated CNT into the Al block, the CNT is placed inside the block and any Al atoms are removed that are located either inside the CNT or within a distance of less than 3.0  $\mathrm{\AA}$ from C  and/or Ni atoms. The system is subsequently relaxed using the FIRE algorithm. A detailed illustration of the resulting Al-CNT composite block with one edge dislocation (split into two Shockley partial dislocations after the minimization) is shown in Fig. \ref{fig:1}. In this figure, $S$ is the distance between the CNT and its image due to the periodic boundary conditions in $z$ direction. It can be calculated as $S = L_z - D$ where $L_z$ is the simulation box length in $z$ direction and $D$ is the CNT diameter.

In the present work, we investigate the influence of three parameters on the critical shear stress ${\tau_c}$ required for the dislocation to pass the array of periodically replicated embedded CNTs. Thus we vary the size of the simulation box in $z$ direction, which in view of the periodic boundary conditions determines the CNT spacing \(S\), as well as the CNT diameter \(D\) and the matrix-CNT interface, which is modified by considering both CNTs and coated NiCNTs. For NiCNTs, $D$ is defined as the outer diameter of the Ni coating. Table \ref{tabl:DSE} details the geometrical parameters used in the simulations.  

\begin{table}
	\centering
	\caption[Characteristic of the simulation of CNT diameter and spacing]{Geometry of the samples used for investigating the influence of CNT diameter and spacing on CRSS. Length unit is $\mathrm{\AA}$.}
	\begin{tabular}{llllll}
		\hline\noalign{\smallskip}
		Samples&$L_x$ &$L_y$& $L_z$& $D$ & $S$ \\
		\noalign{\smallskip}\hline\noalign{\smallskip}
		Pristine CNT&&&&&\\
		1&402&208&99&28&71\\
		2&402&208&198&28&170\\
		3&402&208&297&28&269\\
		4&402&208&396&28&368\\
		5&402&208&382&14&368\\
		6&402&208&438&42&368\\
		Ni-coated CNT&&&&&\\
		1&402&208&396&28&368
	\end{tabular}	
	\label{tabl:DSE}
\end{table} 

The initial configuration of a typical simulation is shown in Fig. \ref{fig:1}. During the relaxation step which follows insertion of the CNT, the dislocation aligns with the CNT and its periodic images, reducing its line energy by intersecting the surface of the hollow cylinder surrounding the CNT. This represents the minimum energy configuration of the unloaded dislocation-CNT system. For later use we introduce the following notations: The origin of the coordinate system is located at the point where the slip plane intersects the centroidal axis of the CNT. The arm of the dislocation extending from the CNT in $+z$ direction to the periodic boundary is denoted as 'upper' arm (UA), the arm extending in $-z$ direction as 'lower' arm (LA). The local dislocation orientation is characterized by the angle $\theta$ between the dislocation line direction and the $z$ direction. Angles are distinguished by subscripts (u,l) where $u$ refers to the upper and $l$ to the lower arm, and by superscripts (1,2) where 1 denotes the leading and 2 the trailing partial dislocation. The points of intersection of the leading (trailing) partial dislocation of the upper (lower) arm with the cylinder surrounding the CNT are accordingly denoted as $I_{u,l}^{1,2}$. In cylindrical coordinates with the CNT axis as cylinder axis these intersection points are located at the points $(R,\phi_{u,l}^{1,2})$ where $R=D/2$, i.e., the angle $\phi_{u,l}^{1,2}$ defines the direction of the vector connecting the CNT center and the intersection point of the leading (trailing) partial of the upper (lower) dislocation.

\subsection{Atomistic simulation}
\label{sec:2.3}
Shear deformation of the simulated sample is imposed by rigidly displacing the top boundary layer in $x$ direction  with a velocity $v_x$ which is constant in time, while a velocity of zero is imposed on the atoms in the lower boundary layer.  Atoms located between the top and bottom boundary layers are assigned initial velocities in $x$ direction according to

\begin{equation}\label{eq:1}
v = v_x c_y/(L_y - 2t_b) 
\end{equation}
where $c_y$ is the distance of the atom from the constrained bottom layer, and $L_y$ and $t_b$ are the simulation box length and the thickness of the top and bottom boundary layers in $y$ direction. This mode of displacement corresponds to a simple shear test. The imposed shear strain rate $\dot{\gamma_{xy}}$ is given by

\begin{equation}\label{eq:2}
\dot{\gamma_{yx}} = v_x/(L_y - 2t_b)  
\end{equation}

The shear strain rate used in our simulations has a constant value of $10^{-6}$ ps$^{-1}$. All shear simulations are performed in a MD framework at finite temperature \(T\) = 0.1 K with $NVT$ ensemble and a timestep size of 1.0 fs. Atomic stresses are calculated based on the virial stress as average virial stresses of all atoms within the specimen volume, excluding the constrained top and bottom layers \cite{Thompson2009-J.chem.phys}.  
 
\section{Results and discussion}
\label{sec3}

\begin{figure}
	\centering
    \includegraphics[width=.8\textwidth]{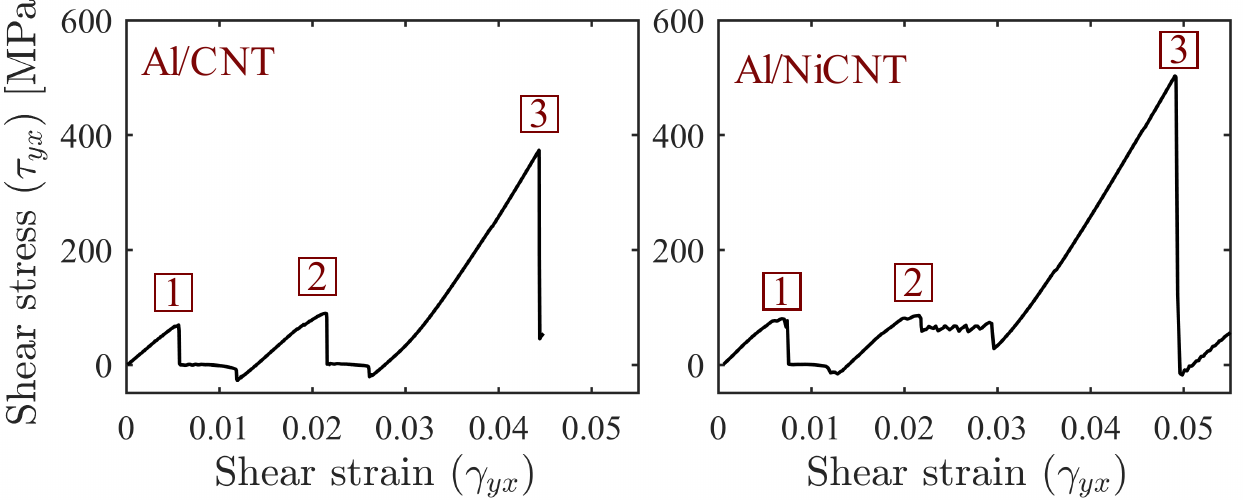}
  \caption[Shear stress-strain curves for Al/(Ni)CNT]{Shear stress-strain curves for Al/CNT (left) and Al/NiCNT (right). The three stress peaks $\fbox{1}$, $\fbox{2}$, $\fbox{3}$ in these curves correspond to the CRSS for the dislocation to pass the obstacle for the first, second, and third time, respectively.}
  \label{figdis:2}    
\end{figure}

During the initial relaxation step, the dislocation reduces its energy by migrating towards the CNT: once the part of the dislocation which intersects the void surrounding the CNT disappears, the dislocation line energy decreases. Upon loading, the imposed shear deformation creates a shear stress in the sample which, in turn, exerts a configurational force (Peach-Koehler-force) that drives the dislocation to move in the positive $x$ direction. Because of the attractive force between the dislocation and the CNT, the dislocation remains first attached to the CNT until, at a critical stress, it breaks free. This process is indicated by a sudden load drop on the stress-strain curves (marked $\fbox{1}$ in Fig. \ref{figdis:2} which shows shear stress-strain curves for Al/CNT and Al/NiCNT systems with $S$ = 368 $\mathrm{\AA}$ and $D$ = 28  $\mathrm{\AA}$). For systems containing many dislocations and CNTs with comparable geometrical arrangement, the stress before this load drop may be interpreted as the critical resolved shear stress (CRSS). 

Due to the periodic boundary conditions imposed in slip direction, a dislocation can move mulitple times through the simulated sample and interact with the embedded CNT. While this appears at first glance unphysical, the situation can be understood as a simulation of multiple dislocations emitted from the same source and sequentially interacting with the CNT.  During these interactions, the conformations of both the dislocation and Al atoms surrounding the CNT, though not of the CNT itself, undergo characteristic changes which lead to characteristic signatures in the stress-strain curves.  The situation may be discussed in analogy with precipitation-hardened materials where for shearable precipitates the cutting of a precipitate by multiple dislocations moving on the same slip plane can lead to softening: the passing stress is reduced as the number of dislocations cutting the precipitates increases, and in some cases the precipitates may even dissolve \cite{hutchinson2009quantifying}. For non-shearable precipitates, on the other hand, the deposition of Orowan loops around the precipitates may lead to strain hardening.  

In the following we first investigate the dependency of the CRSS on CNT diameter and spacing, and then investigate the hardening mechanisms associated with multiple dislocations passing a CNT.

\subsection{Effect of CNT diameter and spacing on CRSS}
\label{sec3.3}

\begin{figure}
\centering
    \includegraphics[width=.95\textwidth]{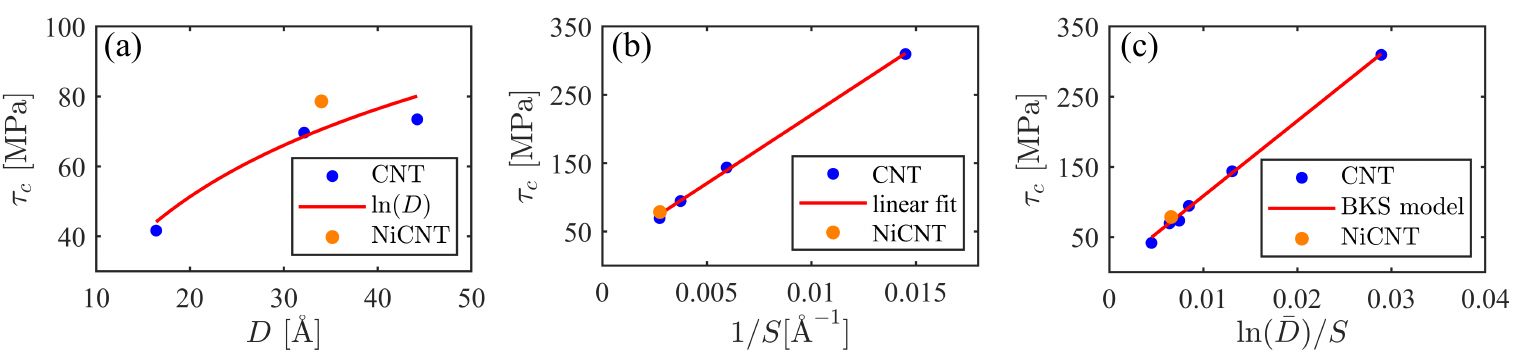} 
  \caption[Effect of $D$ and $S$ on CRSS]{Effects of (a) CNT diameter and (b) CNT spacing on critical resolved shear stress, (c) fit according to BKS model, critical shear stress $\tau_{\mathrm{c}}$ versus $\ln(\bar{D})/S$.}
  \label{fig:dis14}    
\end{figure}

In this part, we investigate the dependency of the CRSS on the CNT diameter $D$ and spacing $S$. Table \ref{tabl:DSE} details the simulation box sizes and CNT diameters. Our MD simulation results show that an increase of the CNT diameter $D$ results in an increase of  CRSS, which can be described by a logarithmic CRSS dependence on $D$, see Fig. \ref{fig:dis14} (a). At the same time, the CRSS changes in approximately inverse proportion with the CNT spacing $S$ as shown in Fig. \ref{fig:dis14} (b). 

Our observations can be well represented by the model proposed by Bacon et al. \cite{Bacon1973-Philos.Mag.} for the interaction of a dislocation with an array of non-shearable precipitates, henceforth referred to as BKS model. This model has been validated for precipitation hardening in metallic alloys by experimental and numerical analysis (\cite{Choi2016-JCOMB, George2005-scripta, Park2018-JMST, Li2009-cops.sci.technol, vaid2019-Materialia}). The model predicts that the critical stress for a dislocation required to pass a periodic array of non-shearable inclusions is given by 

\begin{equation}\label{eq:3}
\tau_{\rm c} = \frac{\mu b}{2 \pi B} ( \frac {\ln(\bar{D})}{S} + C ) 
\end{equation}

where $\mu$ is the shear modulus, ${\bar{D}}$ is the harmonic mean of the spacing $S$ and diameter $D$ of the inclusions, divided by the Burgers vector magnitude $b$: ${\bar{D} = {(b/S + b/D)}^{-1}}$. \(B\) is a nondimensional parameter which depends on dislocation type and is equal to 1 for an edge dislocation, and \(C\) is an adjustable variable. Using values for Al i.e. $b = 2.86 \mathrm{\AA}$ and  $\mu$ = 28 GPa, we find that our MD data are very well represented by the BKS model, see Fig. \ref{fig:dis14} (c).  For an edge dislocation in aluminum gliding in an (111) plane the pre-factor of the above formula is equal to $\frac{\mu b}{2 \pi B}$ = 12.7 (GPa $\mathrm{\AA}$). This value is in good agreement with the slope of a straight line fitted to the MD data of ${\tau_c}$ versus $\frac {\ln(\bar{D})}{S}$, which is 11.4 (GPa). Note that the CRSS values are comparable for coated and uncoated CNTs, provided that the diameter of the Ni coated CNT is taken to be the outer diameter of the coating. 

We thus conclude that, as far as the CRSS is concerned, the embedded CNTs irrespective of coating status behave like non-shearable precipitates. However, when it comes to hardening effects, there are characteristic differences from the classical picture of dislocations by-passing non-shearable obstacles by depositing Orowan loops around them. The hardening mechanisms associated with multiple dislocations passing a CNT on the same slip plane are analyzed in the following.

\subsection{Mechanisms of dislocation-CNT interaction}
\label{sec3.1}

To discuss the detailed processes occurring upon encounter between the moving dislocation and the CNT, we measure the area $A$ swept by the moving dislocation, starting from its equilibrium configuration where it is aligned with the CNT in the $\theta = \phi = 0\degree$ configuration. We measure this area in terms of the parameter $\eta = A/(L_z^2)$, for instance, a dislocation that bows out into a perfect half-circular shape in the $\theta = \phi = 90\degree$ configuration would correspond to $\eta = \pi/8$.  

\subsubsection{Pristine CNT, first encounter}
\label{sec3.1.1}

\begin{figure}
	\centering
    \includegraphics[width=0.9\textwidth]{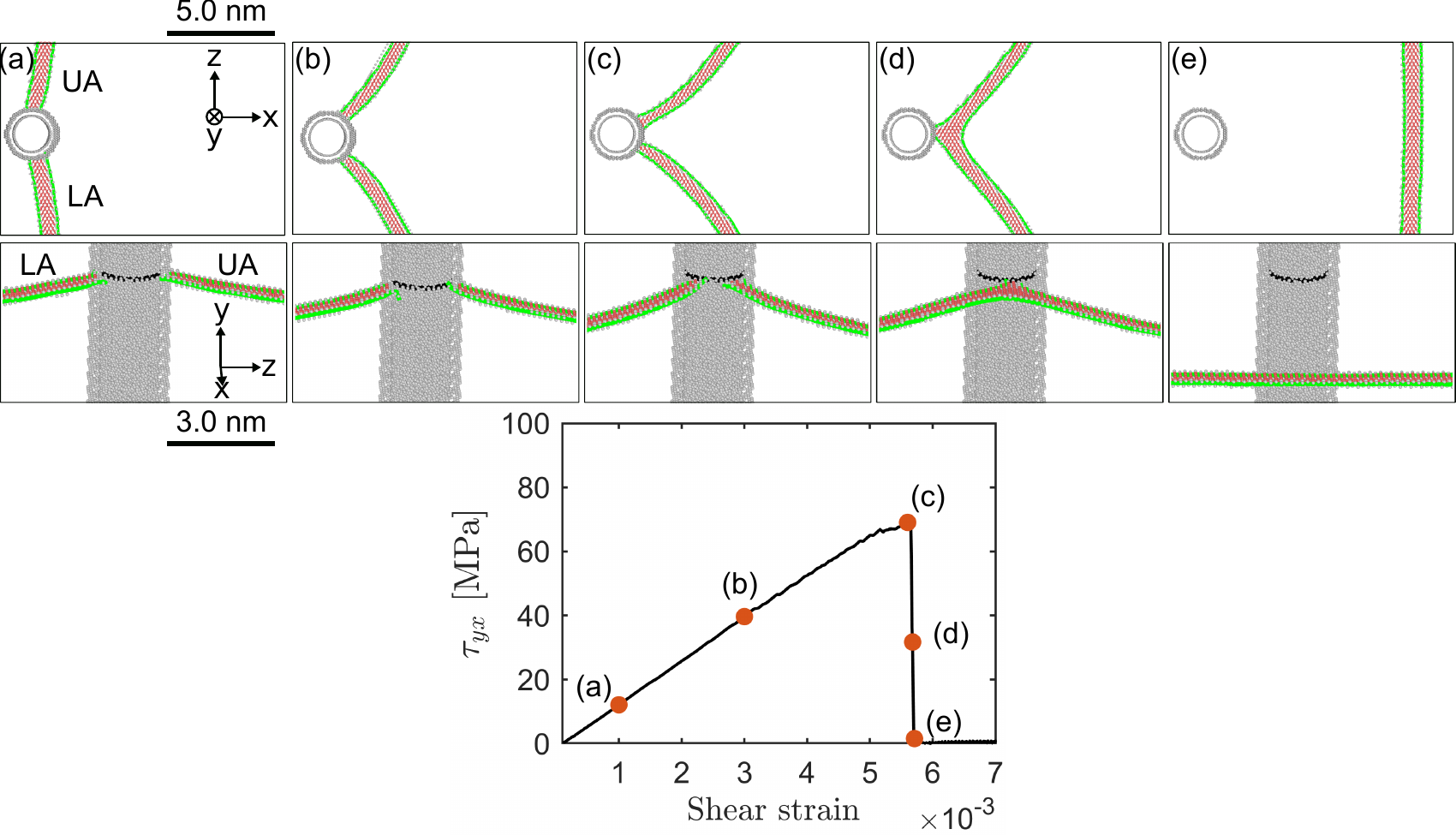}
  \caption[First CNT from top]{Snapshots from the process by which an edge dislocation unpins from a CNT(20,20). (a) $\eta$ = 0.06, (b) $\eta$ = 0.16, (c) $\eta$ = 0.31, (d) $\eta$ = 0.41, and (e) $\eta$ = 0.49. Red atoms represent the stacking fault area between the partial dislocations. The solid green lines indicate partial dislocation lines. Gray atoms shows CNT and unstructured atoms (Al atoms with no crystal structure) around the CNT. In the front views, the trace of the original dislocation slip plane on the CNT is marked in black.} 
  \label{figdis:3}    
\end{figure}

Fig. \ref{figdis:3} shows the first encounter of a dislocation with a pristine CNT(20,20) ($D$ = 28 $\mathrm{\AA}$, $S$ = 368 $\mathrm{\AA}$) in top and front views. In this figure, we only depict atoms with non-fcc crystal environment for clarity. Fig. \ref{figdis:3} (a-d) shows that the dislocation segments on both sides of the CNT move in $x$ direction to accommodate the shear displacement while the points of intersection between the CNT and the dislocation move around the CNT. This motion is concomitant with a rotation of the upper and lower arms of the dislocation. Once the leading partials of both arms have rotated by an angle close to 
$\phi_{u,l}^1 = 90 \degree$ around the CNT, they merge upon encounter (Fig. \ref{figdis:3} (d)). The dislocation then detaches from the CNT, leaving behind a step of height $b \sin \phi$ in the cylindrical Al surface surrounding the void.  This step is accommodated by elastic deformation of the embedded CNT. 

\subsubsection{Second encounter}
\label{sec3.1.2}

\begin{figure}
	\centering
    \includegraphics[width=0.9\textwidth]{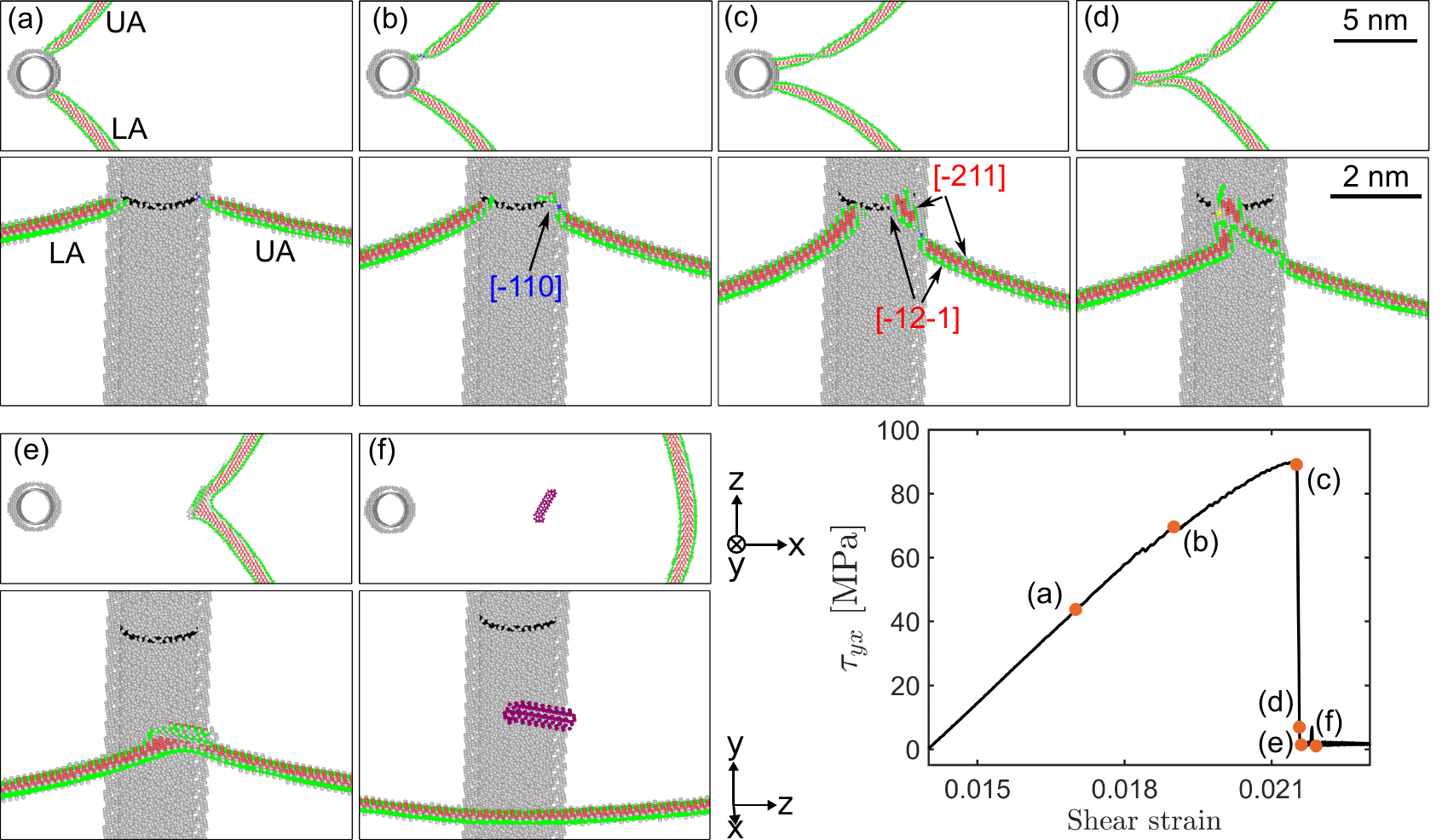}
  \caption[second CNT from top]{Snapshots from the second encounter of dislocation with a CNT(20,20). (a) $\eta$ = 0.14, (b)  $\eta$ = 0.24, (c) $\eta$ = 0.39, (d) $\eta$ = 0.53, (e) $\eta$ = 0.65, and (f) $\eta$ = 0.67. For colour descriptions, refer to Fig. \ref{figdis:3}.} 
  \label{figdis:4}    
\end{figure}

During the second encounter, the dislocation is first attracted towards the CNT and passes a minimum energy configuration where it aligns near-perpendicularly with the CNT in $\theta = \phi = 0\degree$ configuration. Under increasing shear stress, both dislocation arms again start to rotate and the points of intersection migrate around the CNT. However, accomplishing a full quarter circle to the $ \phi^1_{u,l} \approx 90 \degree$ orientation is associated with a prohibitive energy cost since the elastic energy of the created surface step, which must internally be accommodated by a S-shaped bending of the embedded CNT, scales in proportion with the square of the step height. Therefore, the movement of the intersection points is impeded (Fig. \ref{figdis:4} (a)) and the two dislocation arms reach the $\theta_{\rm u,l} = 90 \degree$ screw orientation before the intersection points have reached the $\phi_{\rm u,l} = 90 \degree$ location where the arms could merge. Instead, once the upper intersection point has reached $\phi_{\rm u} \approx 30 \degree$ (Fig. \ref{figdis:4} (b)), the upper arm is already in near-screw orientation. This allows the dislocation to avoid the obstacle presented by the surface step deposited in the previous encounter through a double cross-slip process: the arm constricts at the intersection points, cross slips, and re-dissociates on a parallel slip plane above the original one, thus avoiding the pre-existing slip step (Fig. \ref{figdis:4} (c)). The intersection point of this arm has now acquired increased mobility and travels around the CNT beyond the $\phi_{\rm u}=90 \degree$ location to reach the lower arm (Fig. \ref{figdis:4} (d)), where both arms annihilate via a second cross slip event. This annihilation detaches the dislocation, which in the process has acquired a double jog (Fig. \ref{figdis:4} (e)). This double jog is finally left behind in form of  a prismatic dislocation loop (Fig. \ref{figdis:4} (f)). %

\subsubsection{Third encounter}
\label{sec3.1.3}

\begin{figure}
	\centering
    \includegraphics[width=0.9\textwidth]{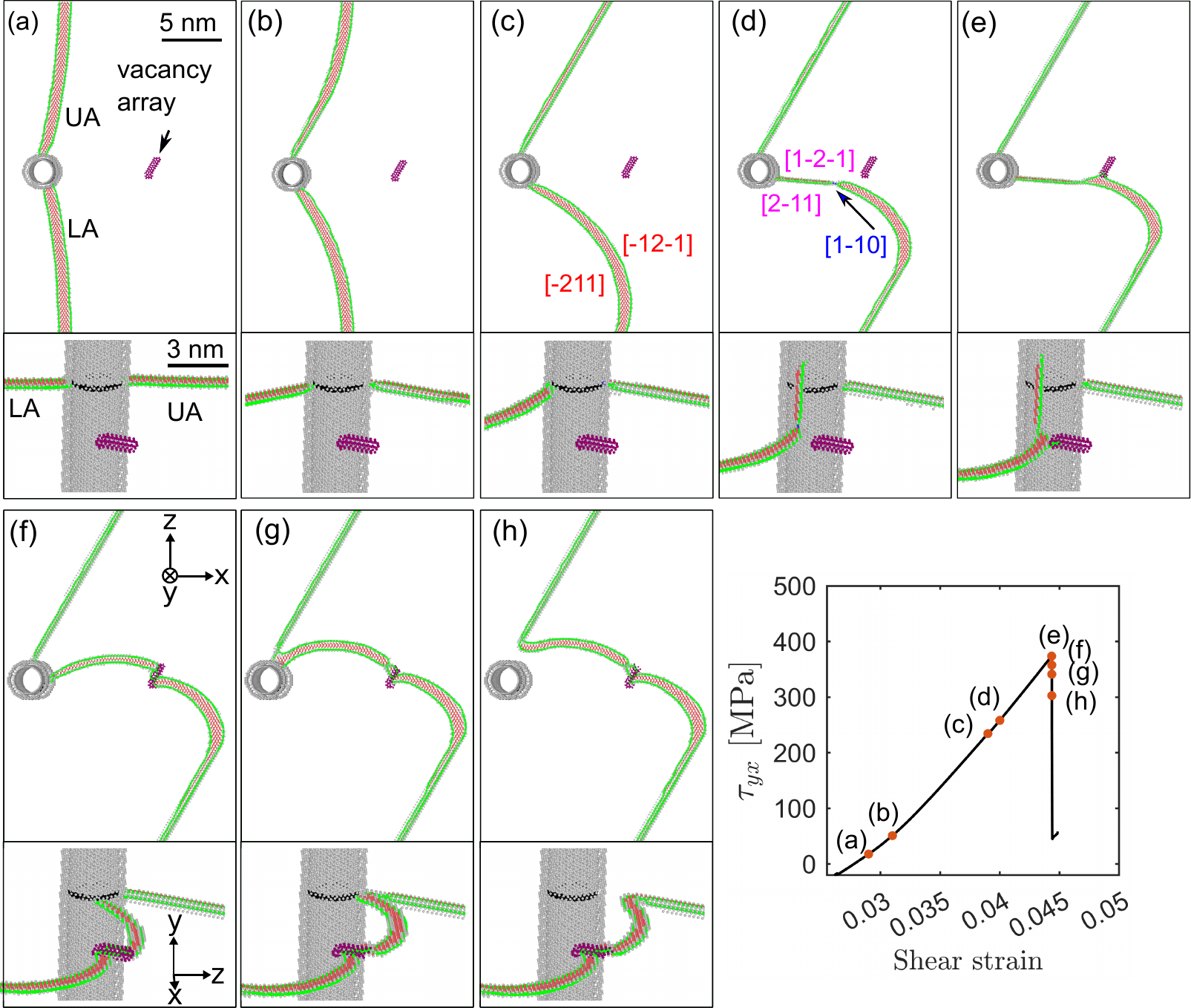}
  \caption[Third CNT]{Snapshots from third encounter of dislocation with a CNT, side view. (a) $\eta$ = 0.05, (b) $\eta$ = 0.13, (c) $\eta$ = 0.34, (d) $\eta$ = 0.36, (e) $\eta$ = 0.44, (f) $\eta$ = 0.47, (g) $\eta$ = 0.51, (h) $\eta$ = 0.59. For colour description, refer to Fig. \ref{figdis:3}.} 
  \label{figdis:7}    
\end{figure}

Fig. \ref{figdis:7} show the process of dislocation unpinning from the CNT for the third time in top and front view. As a result of the previous two encounters, there is now an asymmetry between the upper and lower arms of the dislocation as the intersection between the two slip steps deposited by the upper arm on the CNT-Al interface acts as a very strong pinning point that cannot be passed by the upper arm, which can neither proceed on the original slip plane nor pass onto the cross slip plane (Fig. \ref{figdis:7} (a)). Instead, the partials of the upper arm rotate around their intersection points and, once it reaches the $\theta_{\rm u} = 30 \degree$ orientation, the leading partial -- which is at that moment in screw orientation -- cross slips to create a non-planar core configuration akin to a Lomer-Cottrell lock (Fig. \ref{figdis:7} (b)). The lock then expands along the upper arm and locks it in $\theta_{\rm u} = 30 \degree$ orientation (Fig. \ref{figdis:7} (c)).

The lower arm, on the other hand, undergoes very much the same scenario as the upper arm did in the second encounter: The intersection point moves around the CNT, but its motion is impeded by the previously formed slip step. The arm thus rotates into near screw orientation ($\theta_{\rm l} \approx 90 \degree$) which is reached when the intersection point is located near $\phi_{\rm l} = 45 \degree$ (Fig. \ref{figdis:7} (c)). The lower arm then constricts at the intersection point,
cross slips and re-dissociates on the cross slip plane. This process restores the mobility of the intersection points which move along the intersection of the CNT with the cross slip plane (Fig. \ref{figdis:7} (d,e)). At the same time, the non-cross slipped part of the lower arm continues to rotate until it gets pinned by the prismatic loop that has been left behind from the second dislocation-CNT interaction (Fig. \ref{figdis:7} (e)). The constriction separating the cross slipped and non cross slipped parts of the lower arm also migrates towards the prismatic loop and the interaction causes the cross slipped part to revert to the original slip plane (Fig. \ref{figdis:7} (f)). This part of the lower arm starts rotating around the CNT again, until it eventually meets the upper arm, detaches it from the CNT and unzips the lock (Fig. \ref{figdis:7} (g,h)). 

We note that the process described here is contingent on interaction of the dislocation with the previously deposited prismatic loop and can therefore not occur in the same form when multiple dislocations interact with a CNT in a non-periodic system. While this might lead one to consider the entire process an artefact, we note that a very similar scenario which does {\em not} involve debris from previous passes is observed with Ni-coated CNTs. This will be described in the following. 

The CRSS for the second passage is slightly higher than the first passage. On the other hand, the third passage requires much higher stress as shown in Fig. \ref{figdis:2}. 
\subsection{Dislocation NiCNT interaction}
\label{sec3.2}

\begin{figure}
	\centering
    \includegraphics[width=0.9\textwidth]{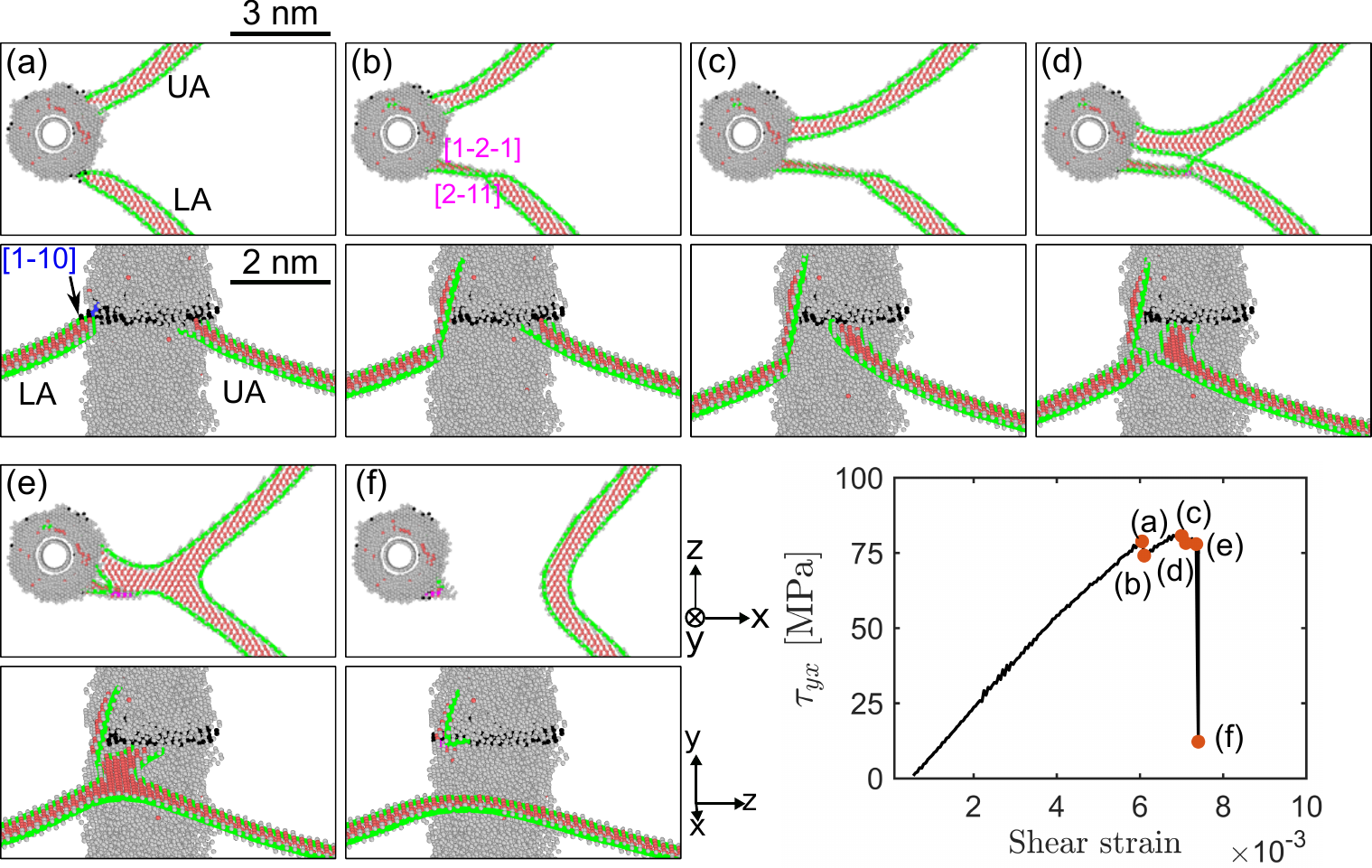}
  \caption[first NiCNT from top]{Snapshots from first encounter of dislocation with a NiCNT. (a) $\eta$ = 0.32, (b) $\eta$ = 0.34, (c) $\eta$ = 0.4, (d) $\eta$ = 0.42, (e) $\eta$ = 0.44, (f) $\eta$ = 0.61. For colour description, refer to Fig. \ref{figdis:3}.} 
  \label{figdis:8}    
\end{figure}

Next, we analyze the dislocation passing mechanisms for a Ni-coated CNT, again considering three sequential CNT-dislocation encounters. The main features of these encounters are similar to the case of a pristine CNT. However, the large elastic mismatch between Al and Ni increases the energy associated with a slip step in the surface of the cylinder surrounding the CNT interface, which now is doubled up by a step in the interface between Al matrix and Ni coating. As a consequence, migration of the dislocation intersection point around the CNT is impeded already during the first encounter. Thus, the lower arm reaches the $\theta_{l} = 90 \degree$ orientation, constricts and passes on the  cross slip plane when its intersection point with the Ni coating is still at $\phi_{\rm l} \approx 45 \degree$. The cross slipped part then expands on the cross-slip plane while the constriction point travels along the intersection of the primary and cross slip planes. At the same time, the attractive interaction between upper and lower arms helps the upper arm to travel around the CNT until it reaches the non-cross slipped part of the lower arm and annihilates (Fig. \ref{figdis:8} (e)). This process detaches the dislocation from the CNT, leaving behind a non-planar half dislocation loop which is deposited at the NiCNT-Al interface.  

\begin{figure}
	\centering
    \includegraphics[width=0.9\textwidth]{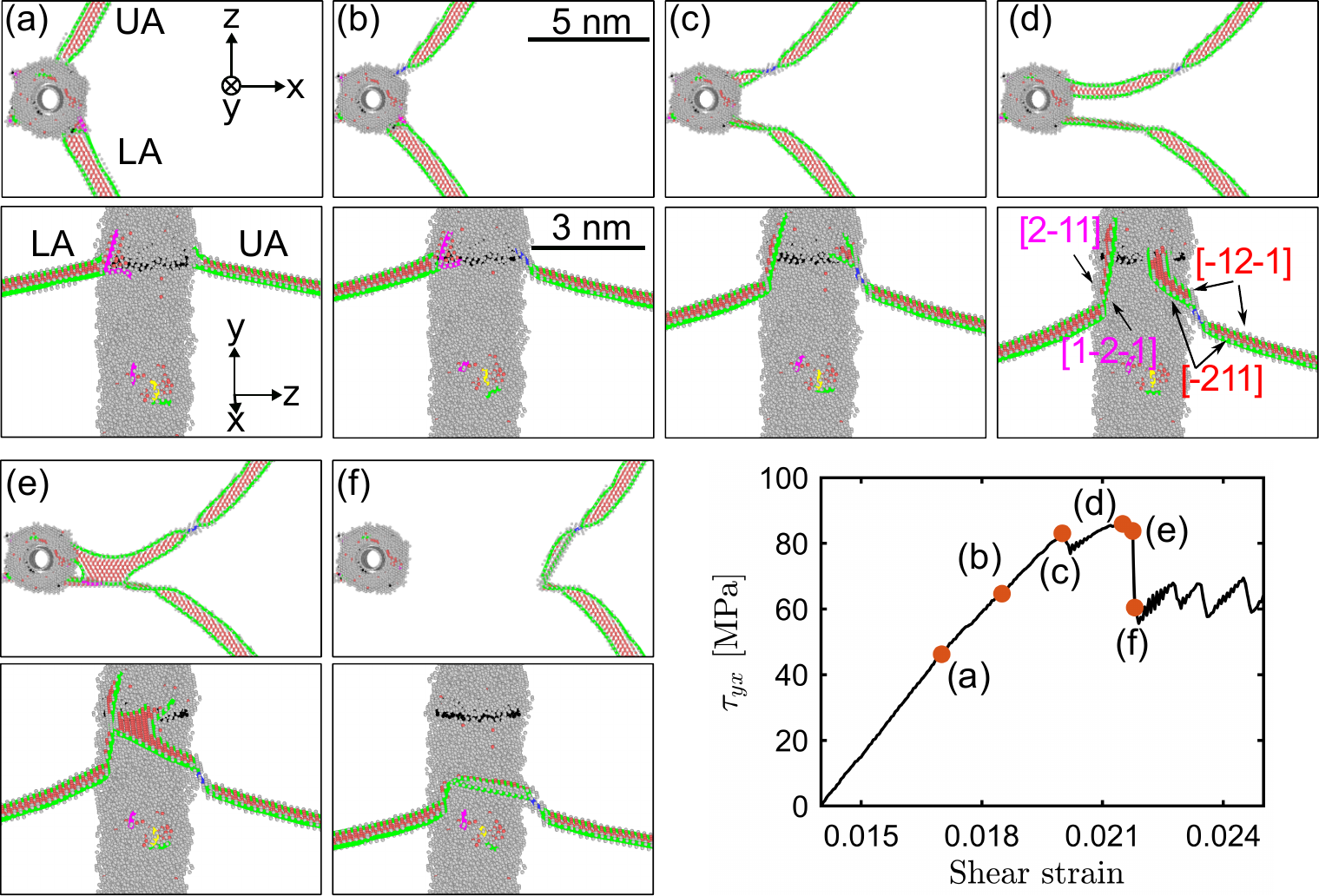}
  \caption[second NiCNT from top]{Snapshots from second encounter of dislocation with a NiCNT. (a) $\eta$ = 0.13, (b) $\eta$ = 0.21, (c) $\eta$ = 0.29, (d) $\eta$ = 0.42, (e) $\eta$ = 0.45, (f) $\eta$ = 0.52. For colour description, refer to Fig. \ref{figdis:3}.} 
  \label{figdis:10}    
\end{figure}

During the second encounter between NiCNT and dislocation the lower arm again passes on a cross slip plane  (Fig. \ref{figdis:10} (a)). However, the motion of the upper arm is modified since now the pre-existing slip step prevents its motion around the CNT on the original slip plane. Instead, the upper arm undergoes a {\rm non-crystallographic} double cross slip event (Fig. \ref{figdis:10} (b,c)) which takes it to a parallel slip plane. The double-cross-slipped segment then travels around the CNT (Fig. \ref{figdis:10} (d)) until it reaches the cross-slipped segment of the lower arm and the mutual attraction of both segments causes the latter to reverse its motion and annihilate (Fig. \ref{figdis:10} (e)), thus detaching the dislocation (Fig. \ref{figdis:10} (f)). In the process the dislocation again acquires a double jog.  The stress-strain curve of the AlNiCNT system shows (Fig. \ref{figdis:2}) that despite the unpinning from the NiCNT, the shear stress does not drop significantly. This is because of the presence of the super jog on the dislocation, which requires much energy to glide. Eventually, the dislocation leaves behind the double jog in form of a prismatic dislocation loop and reverts to a planar configuration. 

\begin{figure}
	\centering
    \includegraphics[width=0.9\textwidth]{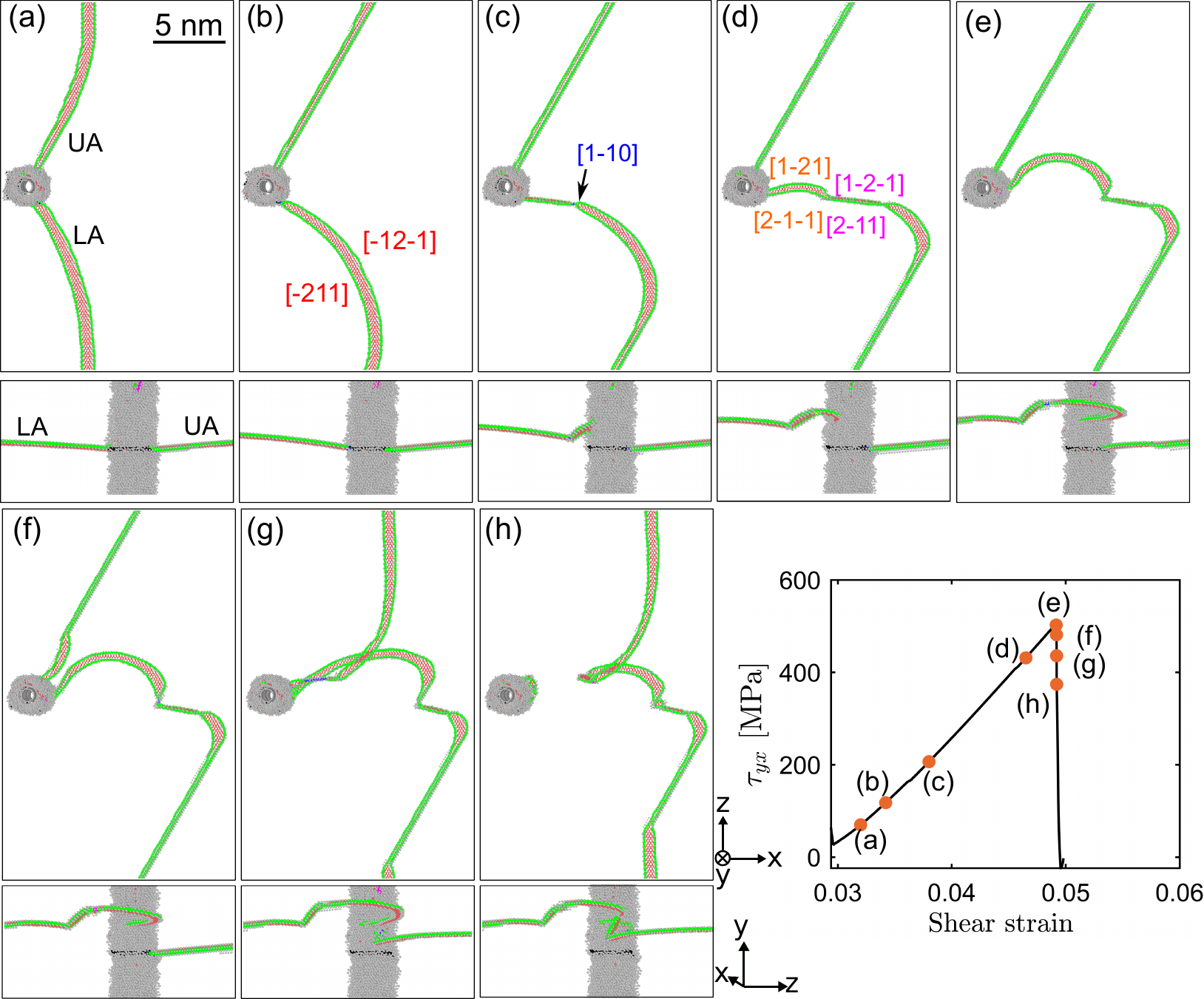}
  \caption[third NiCNT from top]{Snapshots from third encounter of dislocation with a NiCNT. (a) $\eta$ = 0.03, (b) $\eta$ = 0.11, (c) $\eta$ = 0.22, (d) $\eta$ = 0.4, (e) $\eta$ = 0.46, (f) $\eta$ = 0.51, (g) $\eta$ = 0.63, (h) $\eta$ = 0.69}
  \label{figdis:12}    
\end{figure}

The scenario during the third encounter with the NiCNT closely resembles the findings in case of a pristine CNT: The upper arm -- now strongly pinned by the slip step and debris left behind from the previous encounters -- undergoes cross slip of the leading partial once it reaches the $\phi_{\rm u} = \theta_{\rm u} = 30 \degree$ configuration and as a consequence transforms into a lock
(Fig. \ref{figdis:12} (a,b)).  The lower arm constricts at the intersection point with the Ni coating and expands on its cross-slip plane while the constriction point travels along the intersection line of primary and cross slip plane (Fig. \ref{figdis:12} (c)). Ultimately, the cross slipped segment undergoes a second cross slip event and passes onto a slip plane parallel to the primary one (Fig. \ref{figdis:12} (c,d)), where it rotates around the Ni-coated CNT (Fig. \ref{figdis:12} (e)) and ultimately approaches the locked upper arm. The repulsive interaction between the leading partials of upper and lower arms then causes the leading partial of the upper arm to revert to its original glide plane, the lock becomes unzipped and the upper arm is re-mobilized (Fig. \ref{figdis:12} (f)). Finally, a cross slip event of the upper arm (Fig. \ref{figdis:12} (g)) causes the upper and lower arms to merge and detach from the CNT (Fig. \ref{figdis:12} (h)), leaving the dislocation  with a double super-jog. We note that this process is generic in the sense that it does not involve any dislocation debris.

\section{Discussion and conclusions}

We have studied the interactions of edge dislocations with pristine and Ni-coated CNT in Al matrices. These obstacles are found to behave in a manner that is, in a sense, intermediate between non-shearable and shearable precipitates. While the CRSS dependency on CNT diameter and spacing is well described by relations derived originally for non-shearable precipitates, the details of the interaction mechanism differ significantly from a simple Orowan mechanism. This is particularly true for multiple dislocations passing the obstacle on the same slip plane. While the CNT in itself is not shearable, the cylindrical void surrounding the CNT can be sheared at the expense of forming an interfacial slip step in conjunction with elastic deformation of the CNT. This configuration acts as an obstacle to subsequent dislocations -- an effect that is enhanced by Ni coating of the CNT. Second and third dislocations passing the CNT on the same slip plane therefore require increasing shear stresses to overcome the obstacle, i.e., the sequential interaction of multiple dislocations with an embedded CNT leads to strain hardening, quite unlike precipitate shearing which may lead to softening or even dissolution of the precipitates. However, the hardening scenario is also quite different from the classical Orowan scenario which envisages the deposition of multiple dislocation loops around non-shearable precipitates. The scenario observed here is much more complex and involves crystallographic and non-crystallographic cross slip of near-screw segments on the CNT-metal interface, dislocation lock formation and annihilation, jog formation and shedding of prismatic dislocation loops. We note, however, that elements of this scenario have been observed in other MD simulations of dislocation-obstacle interactions, see e.g. the observation of double jog formation during passing of a hard precipitate by an edge dislocation in Ni as reported in \cite{Proville2010-Acta}, or the formation of jogs during interaction of edge dislocations in Fe with voids and He bubbles as reported in \cite{hafez2010influence}.

\section*{Acknowledgments}
SN and MZ acknowledge financial support from DFG under grant no. Za171/11-1. The authors would like to thank Erik Bitzek, Julien Gu{\'e}nol{\'e} and Aviral Vaid for discussion and insights regarding atomistic simulation of dislocations in metals. The authors gratefully acknowledge the compute resources and support provided by the Erlangen Regional Computing Center (RRZE).


\section*{Conflict of Interest}
The authors declare that they have no known competing financial interests or personal relationships that could have appeared to influence the work reported in this paper.

\bibliographystyle{elsarticle-num} 
\bibliography{dislocation_AlCNT_arXiv.bib}

\end{document}